\documentclass{iau}
\usepackage{graphicx}
\usepackage{multirow}
\usepackage{natbib}
\bibpunct{(}{)}{;}{a}{}{,}

\title[Scaling laws to understand tidal dissipation in fluids] 
{Scaling laws to understand tidal dissipation in fluid planetary layers and stars}

\author[Pierre Auclair-Desrotour, St{\'e}phane Mathis \& Christophe Le Poncin-Lafitte]   
{Pierre Auclair-Desrotour$^{1,2}$,
St{\'e}phane Mathis$^{2,3}$\\
 \and Christophe Le Poncin-Lafitte$^4$}

\affiliation{
$^1$IMCCE, Observatoire de Paris, CNRS UMR 8028, \\ 77 Avenue Denfert-Rochereau, 75014 Paris, France \\ 
email: {\tt pauclair-desrotour@imcce.fr} \\[\affilskip]
$^2$Laboratoire AIM Paris-Saclay, CEA/DSM - CNRS - Universit\'e Paris Diderot, \\ IRFU/SAp Centre de Saclay, F-91191 Gif-sur-Yvette, France \\ email: {\tt stephane.mathis@cea.fr} \\[\affilskip]
$^3$LESIA, Observatoire de Paris, CNRS UMR 8109, UPMC, Univ. Paris-Diderot,\\ 5 place Jules Janssen, 92195 Meudon, France \\[\affilskip]
$^4$SYRTE, Observatoire de Paris, CNRS UMR 8630, UPMC, \\ 61 Avenue de l'Observatoire, 75014 Paris, France\\ 
email: {\tt christophe.leponcin@obspm.fr}
}

\pubyear{2014}
\volume{310}  
\pagerange{119--126}
\jname{Complex Planetary Systems}
\editors{A.C. Editor, B.D. Editor \& C.E. Editor, eds.}
\begin{document}

\maketitle

\begin{abstract}

Tidal dissipation is known as one of the main drivers of the secular evolution of planetary systems. It directly results from dissipative mechanisms that occur in planets and stars' interiors and strongly depends on the structure and dynamics of the bodies. This work focuses on the mechanism of viscous friction in stars and planetary layers. A local model is used to study tidal dissipation. It provides general scaling laws that give a qualitative overview of the different possible behaviors of fluid tidal waves. Furthermore, it highlights the sensitivity of dissipation to the tidal frequency and the roles played by the internal parameters of the fluid such as rotation, stratification, viscosity and thermal diffusivity that will impact the spins/orbital architecture in planetary systems.


\end{abstract}

\firstsection 
              
\section{Introduction}  

The bodies that compose a planetary system often present one or several fluid regions. For example, stars are completely fluid, gaseous giant planets are mainly constituted of large fluid envelopes, and some rocky planets and icy satellites such as the Earth, Europa or Enceladus have internal and external fluid layers. Submitted a tidal perturbation, these fluid regions move and deform periodically like solid ones. These tidal motions are affected by different kinds of dissipation mechanisms: the viscous friction, thermal diffusion and Ohmic dissipation in the presence of a magnetic field. Theoretical studies of these processes have emerged after the first study of a tidally deformed body performed by Lord Kelvin \citep{Kelvin1863} and the developments made in 1911 by Love, who introduced the so-called Love-numbers \citep{Love1911}. In the 1960's, Goldreich established the link that bounds the long-term evolution of planetary systems to the internal dissipation \citep[see][]{GS1966}. He introduced the commonly used $Q $ tidal quality factor that intervenes in the dynamical equations. Then, several works, such as \cite{EL2007} for rocky bodies, illustrate the impact of dissipation on the evolution of spin and orbital parameters. 

Over the past decades, studies have been carried out on the tidal dissipation in fluid bodies, these studies being mainly about stars and the envelopes of gaseous giant planets \citep{OL2004,OL2007,RMZ2012}. Given the great diversity of existing tidal motions and dissipation mechanisms, the nature of the response varies, taking very different possible complex forms. For example, the internal fluid layers of a planet and the interiors of stars dissipate energy through the mechanism of viscous friction, but their physical parameters vary over several orders of magnitude. This is the reason why a local approach, described further (and fully developed in Auclair-Desrotour, Mathis \& Le Poncin Lafitte, in preparation), is privileged here: it allows to explore the physics using simplified tractable models \citep[see also][]{O2005,J2014} and comes as a complement to global models by providing qualitative results. 

In a fluid, a perturbation can generate a large spectrum of waves. The acoustic waves are the first kind but their characteristic frequency is too high to allow a significative response to a low-frequency tidal perturbation. This later will rather excite inertial waves and gravity waves, the combination of the two kinds being gravito-inertial waves. Inertial waves have the Coriolis force as restoring force. Their typical frequency is the inertial frequency, $ 2 \Omega $, where $ \Omega $ is the spin frequency of the body. Gravity waves, which have the buoyancy as restoring force, propagate in stably stratified fluid layers. They are characterized by the Brunt-V{\"a}is{\"a}l{\"a} frequency, often denoted $ N $ in the literature, which results from radial variations of the specific entropy. 

These physical parameters, $ \Omega $ and $ N $, like the viscosity $ \nu $ and the thermal diffusivity $ \kappa $ of the fluid have a real impact on the tidal kinetic energy dissipated by viscous friction. The goal of this study is to understand this dissipation. Thus, through the simplified model of a local fluid box, scaling laws are obtained, characterizing the dissipation spectra, to give simplified models that can be used as a first step in celestial mechanics studies. In this way, it is demonstrated that dissipation strongly depends on the forcing frequency and internal parameters in the case of fluids, and so do the spin and orbital evolution of corresponding planetary system \citep{ADLPM2014}.

\begin{figure}[htb]
 \centering
 \includegraphics[width=0.46\textwidth,clip]{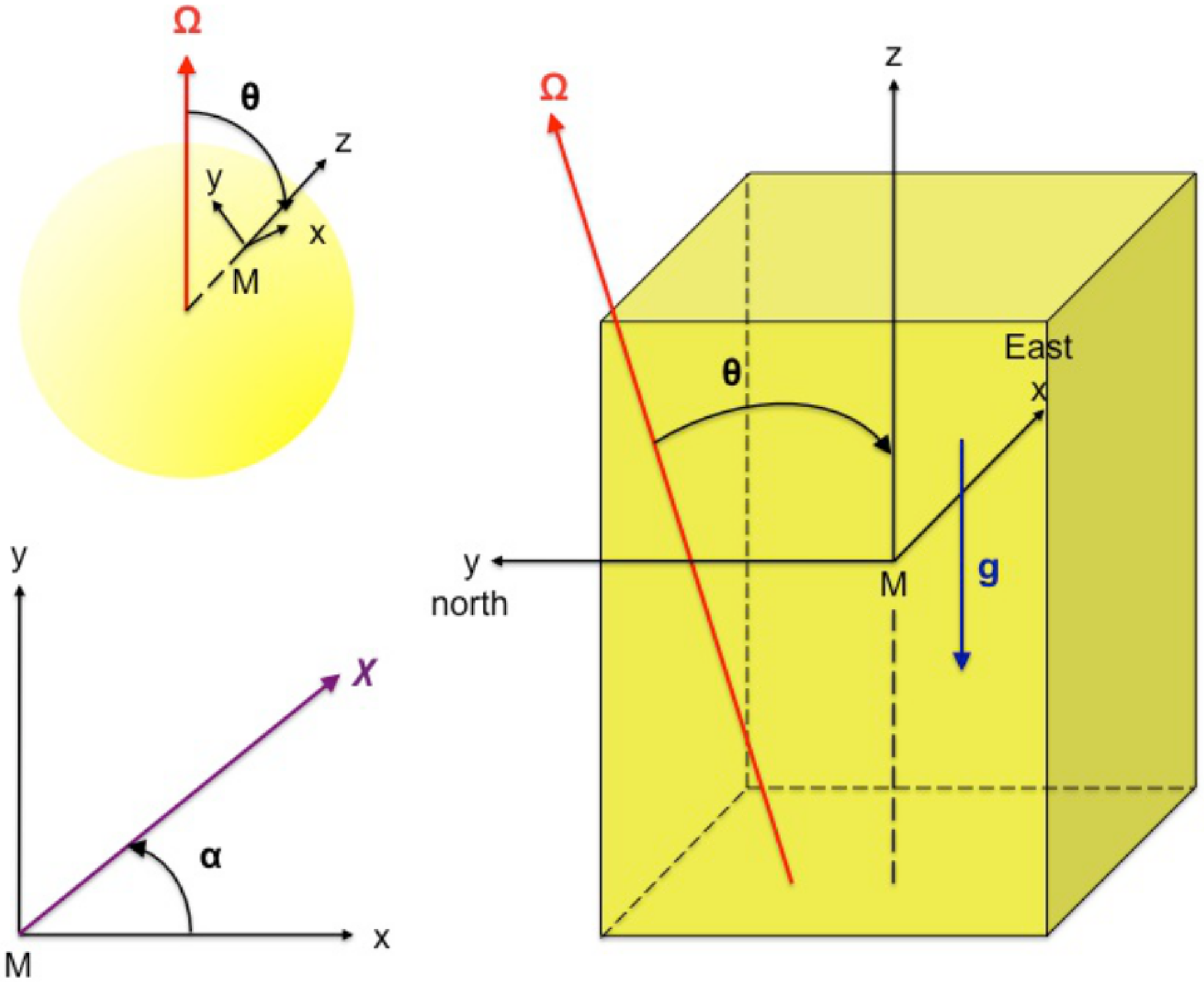}      
 \includegraphics[width=0.46\textwidth,clip]{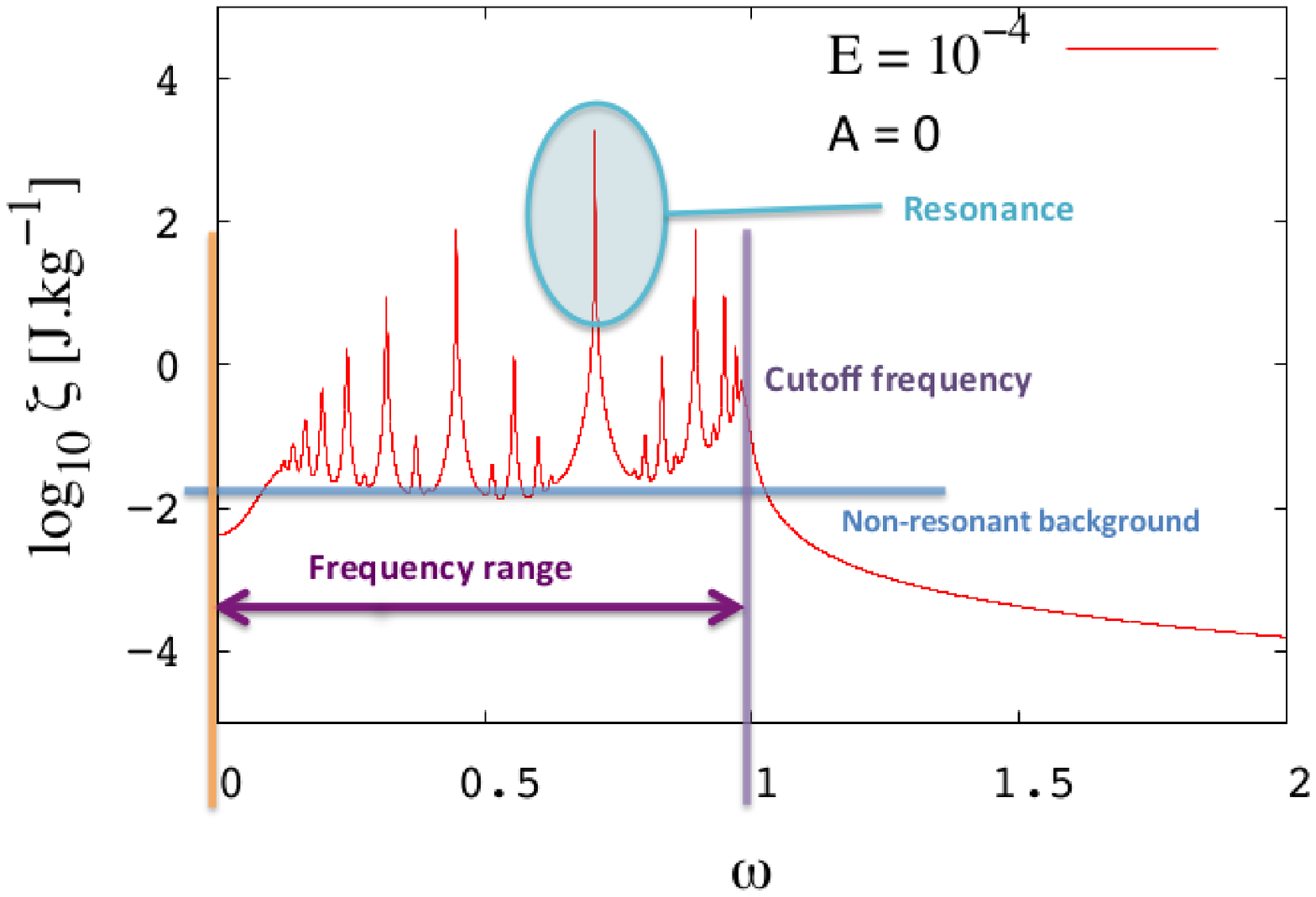}  
  \caption{\label{fig:box}  {\bf Left:} The studied model: an inclined rotating Cartesian fluid box where the viscosity, thermal diffusion, and stratification of the fluid are taken into account. {\bf Right:} A typical dissipation frequency spectrum computed from the analytical model. $ \zeta $ is the mass energy dissipated by viscous friction per rotation period in the box ; $ \omega = \chi / 2 \Omega $ is the tidal frequency of the perturbation normalized by the inertial frequency $ 2 \Omega $. The Ekman number ($E$) and the stratification parameter ($A$) are defined in the text hereafter.}
\end{figure}

\section{Model and results: waves in a box}

Consider a local section of a fluid region in any star or planet. It can be represented as a rotating Cartesian box of side length $ L $ inclined with respect to the spin axis of the body $ \boldsymbol{\Omega} $ with a colatitude $ \theta $ (Fig. \ref{fig:box}). This model generalizes the first local model presented in \cite{OL2004}. The local coordinates $ \left( x,y,z \right) $ are defined so that $ z $ corresponds to the radial direction, and $ x $ and $ y $ to the West-East and South-North directions respectively. The fluid is supposed to be locally homogeneous of background density $ \rho $, kinetic viscosity $ \nu $, and thermal diffusivity $ \kappa $. This implies three control parameters, denoted $ A $, $ E $, and $ K $. The behavior of the waves is given by $ A = \left( N / 2 \Omega \right)^2 $: inertial waves correspond to $ A \ll 1 $, and pure gravity waves to $ A \gg 1 $. Then, the so-called Ekman number $ E = 2 \pi^2 \nu / \left(\Omega L^2\right) $, and $ K = 2 \pi^2 \kappa / \left(\Omega L^2\right) $ weight the effects of viscous and thermal diffusions with respect to the Coriolis effect. Their ratio forms the Prandlt number $ Pr = E/K $. The fluid is submitted to a tidal perturbative force of frequency $ \chi $.

\begin{figure}[htb]
\begin{center}
 \includegraphics[width=0.825\textwidth]{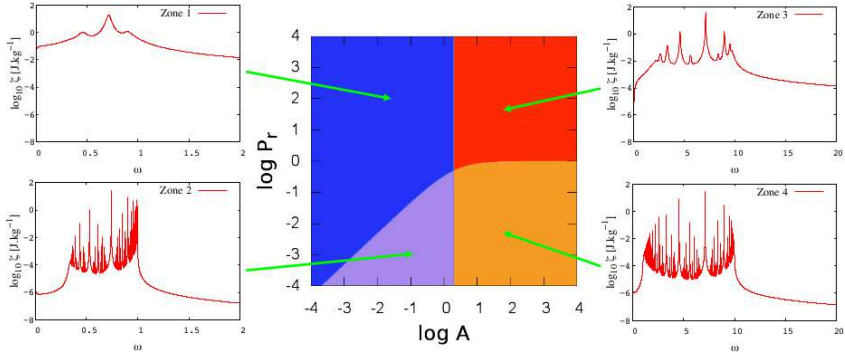} 
 \caption{Asymptotical behaviors of the tidal waves. Zones colored in blue and purple correspond to inertial waves, the two other to gravity waves ; zones colored in blue and red correspond to the case where viscous diffusion predominates over thermal diffusion, the two zones below corresponding to the opposite case.}
   \label{fig:map}
\end{center}
\end{figure}

\begin{figure}[htb]
\begin{center}
 \includegraphics[width=0.7\textwidth]{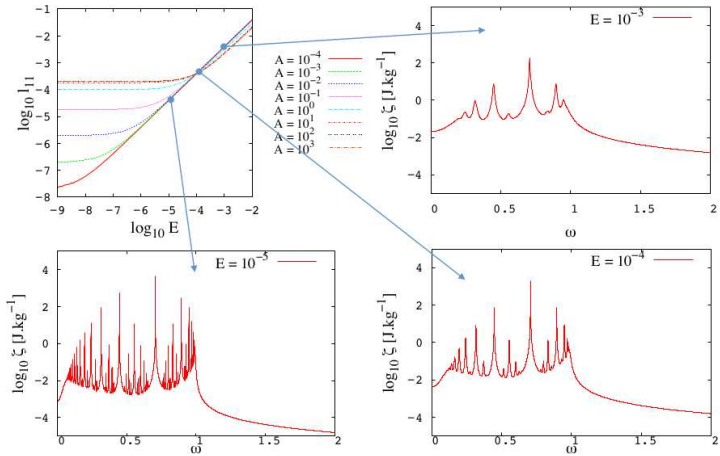} 
 \caption{Width at mid-height of the main resonance as a function of $ E $ for different values of $ A $ and the corresponding dissipation spectra.}
   \label{fig:largeur}
\end{center}
\end{figure}

The energy dissipated by viscous friction in the box is computed analytically. Expressed as a function of $ \omega = \chi / 2 \Omega $, the normalized frequency of the perturbation, it yields the dissipation spectrum of the system (Figs. \ref{fig:box}, \ref{fig:map} and \ref{fig:largeur}). Note that this latter takes the form of a batch of resonances located between $ \chi_1 \approx 2 \Omega \cos \theta $ and $ \chi_2 \approx N $. Its properties, such as the positions $ \omega_{mn} $, widths $ l_{mn} $ and heights $ H_{mn} $ of the peaks, $ m $ and $ n $ being respectively the horizontal and vertical wave numbers, the number of resonances $ N_{ \rm kc} $, the height of the non-resonant background $ H_{\rm bg} $ (corresponding to the equilibrium tide) and the sharpness ratio $ \Xi = H_{11} / H_{\rm bg} $, depend on the regime of the waves highlighted by the model (Fig. \ref{fig:map}). For each regimes, these properties are described by scaling laws obtained analytically as functions of the control parameters and synthesized in Table \ref{tab1}. An illustrative example is given in Fig. \ref{fig:largeur} for $l_{11}$.

\section{Conclusions}            

The analytical local model of the fluid box appears as a useful and efficient way to explore a large domain of physical parameters. It allows to identify and to describe the asymptotical behaviors of the fluid qualitatively and can be used in parallel with more complex global models to understand the physics of dissipation that must be used in celestial mechanics' studies. In this study, it has been restricted to viscous friction but it is possible to take the other dissipations, enumerated in introduction, into account. One could also think about applying it in a near future to Alfv{\'e}n waves, and not only to gravito-inertial waves, by adding a magnetic field to the model, stars and planets being magnetized bodies.

\begin{table}[htb]
\centering
  \caption{Scaling laws of the properties of the energy viscously dissipated for the different regimes (we define $A_{11}\equiv2\cos^{2}\theta$ and $Pr_{11}\equiv A/\left(A+A_{11}\right)$). {\bf Top left:} Inertial waves dominated by viscosity. {\bf Top right:} Gravity waves dominated by viscosity. {\bf Bottom left:} Inertial waves dominated by heat diffusion. {\bf Bottom right:} Gravity waves dominated by heat diffusion. $ F $ is the amplitude of the forcing.}
  \vspace{1mm}
  \label{tab1}
{\scriptsize
\begin{tabular}{ c | l l | l l}
  \hline
  \hline
  \vspace{0.01mm} & & & &\\
    \textsc{Domain} & $ A \ll A_{11} $ & & $ A \gg A_{11} $ &  \\ \hline
     \vspace{0.01mm} & & & &\\
  \multirow{4}{*}{$ Pr \gg Pr_{11} $}  & $ \displaystyle \frac{\chi_{mn}}{2 \Omega} \propto \displaystyle \frac{n}{\sqrt{m^2 + n^2}} \cos \theta $ & $ N_{\rm kc} \propto E^{-1/2} $ &  $ \displaystyle \frac{\chi_{mn}}{2 \Omega} \propto \displaystyle \frac{m}{\sqrt{m^2 + n^2}} \sqrt{A} $ & $ N_{\rm kc} \propto A^{1/4} E^{-1/2}$ \\
     \vspace{0.01mm} & & & & \\
    & $ l_{mn} \propto E  $  & $ H_{mn} \propto F^2 E^{-1} $  & $ l_{mn} \propto E  $ & $ H_{mn} \propto F^2 E^{-1} $ \\
     \vspace{0.01mm} & & & & \\
    & $ H_{\rm bg} \propto F^{2} E $ & $ \Xi \propto E^{-2} $ & $ H_{\rm bg} \propto F^{2} E A^{-1} $ & $ \Xi \propto A E^{-2} $ \\
   \vspace{0.01mm} & & & & \\
     \hline
     \vspace{0.01mm} & & & &\\
  \multirow{3}{*}{$ Pr \ll Pr_{11} $}  & $ \displaystyle \frac{\chi_{mn}}{2 \Omega} \propto \displaystyle \frac{n}{\sqrt{m^2 + n^2}} \cos \theta $  & $ N_{\rm kc} \propto A^{-1/2} K^{-1/2} $  & $ \displaystyle \frac{\chi_{mn}}{2 \Omega} \propto \displaystyle \frac{m}{\sqrt{m^2 + n^2}} \sqrt{A} $ & $ N_{\rm kc} \propto A^{1/4} K^{-1/2} $ \\
       \vspace{0.01mm} & & & & \\
    & $ l_{mn} \propto AK $ & $ H_{mn} \propto F^2 A^{-2} E K^{-2} $ & $ l_{mn} \propto K $ & $ H_{mn} \propto F^2 E K^{-2} $ \\
       \vspace{0.01mm} & & & & \\
    & $ H_{\rm bg} \propto F^{2} E  $ & $ \Xi \propto A^{-2} K^{-2} $ & $ H_{\rm bg} \propto F^{2} E A^{-1} $ & $ \Xi \propto A K^{-2} $ \\
      \vspace{0.01mm} & & & & \\
  \hline
  \hline
\end{tabular}
}
\end{table}

\bibliographystyle{iau310}  
\bibliography{IAU310_auclair-desrotour} 

\end{document}